 \documentclass[final,5p]{elsarticle}
\usepackage{amssymb}
\usepackage{amsmath, nccmath}
\usepackage{lipsum}
\usepackage[normalem]{ulem}
\usepackage{color}
\usepackage{float}

\usepackage{lineno}
\usepackage{hyperref}
\usepackage[shortcuts]{extdash}
\usepackage{siunitx}

\renewcommand{\epsilon}{\varepsilon}
\newcommand{\ur}[1]{$^{23{#1}}$U}
\newcommand{\pu}[1]{$^{2{#1}}$Pu}
\newcommand{\fth}{5$^{\mathrm{th}}$}
\def\antiparticle{\overline}
\def\antinu{$\overline{\nu_e}$}

\journal{Physics Letters B}

\begin{document}

\begin{frontmatter}



\title{Long term remote  reactor power and fuel composition monitoring using antineutrinos}


%\author[first]{Author name}
%\affiliation[first]{organization={University of the Moon},%Department and Organization
%            city={Earth},

\author[a,b,c,d]{I.~Alekseev}
\author[d,e]{V.~Belov}
\author[b,e,f]{A.~Bystryakov}
\author[b,d]{M.~Danilov}
\author[e]{D.~Filosofov}
\author[e]{M.~Fomina}
\author[c]{P.~Gorovtsov}
\author[a,g]{Ye.~Iusko}
\author[d,e]{S.~Kazartsev}
\author[h]{V.~Khvatov}
\author[h]{S.~Kiselev}
\author[a,b,c]{A.~Kobyakin}
\author[a,b,c]{A.~Krapiva}
\author[e]{A.~Kuznetsov}
\author[i]{I.~Machikhiliyan}
\author[a,b,c]{N.~Mashin}
\author[e]{D.~Medvedev}
\author[a,b]{V.~Nesterov}
\author[b,d,e]{D.~Ponomarev}
\author[e]{I.~Rozova}
\author[a,e,f]{N.~Rumyantseva}
\author[a,b]{V.~Rusinov}
\author[e]{A.~Salamatin}
\author[a,b]{E.~Samigullin}
\author[e]{Ye.~Shevchik}
\author[d,e]{M.~Shirchenko}
\author[j]{Yu.~Shitov}
\author[a,b,d]{N.~Skrobova\corref{correspondingauthor}}
\cortext[correspondingauthor]{Corresponding author}
\ead{skrobova@lebedev.ru}
\author[a,b,d]{D.~Svirida}
\author[a]{E.~Tarkovsky}
\author[c]{A.~Yakovleva}
\author[e]{E.~Yakushev}
\author[d,e]{I.~Zhitnikov}
\author[e,k]{D.~Zinatulina}

\affiliation[a]{organization={National Research Center "Kurchatov Institute"}, addressline={Akademik Kurchatov square 1, Moscow, 123182, Russia}}
\affiliation[b]{organization={Lebedev Physical Institute of the Russian Academy of Sciences}, addressline={Leninskiy avenue 53, Moscow, 119991, Russia}}
\affiliation[c]{Moscow Institute of Physics and Technology, addressline={Institutskiy lane 9, Dolgoprudny, Moscow Region, 141701, Russia}}

\affiliation[d]{organization={Institute for Nuclear Research of the Russian Academy of Sciences}, addressline={60th October Anniversary Prospect 7a, Moscow 117312, Russia}}

\affiliation[e]{organization={Joint Institute for Nuclear Research},  addressline={Joliot-Curie str. 6, Dubna, Moscow region, 141980, Russia}}

\affiliation[f]{organization={Dubna State University}, Universitetskaya str. 19, Dubna, Moscow region, 141982, Russia}

\affiliation[g]{organization={National Research Nuclear University Moscow Engineering Physics Institute}, addressline={Kashirskoe shosse 31, Moscow, 115409, Russia}}

\affiliation[h]{organization={JSC Rosenergoatom Concern Affiliate—Kalinin Nuclear Power Plant}, addressline={Udomlya, 171841 Russia}}
\affiliation[i]{organization={Federal State Unitary Enterprise Dukhov Automatics Research Institute}, addressline={Sushchevskaya str. 22, Moscow, 127055, Russia}}

\affiliation[j]{organization={Institute of Experimental and Applied Physics, Czech Technical University in Prague}, addressline={Husova 240$/$5, Prague, 110 00 Czech Republic}}

\affiliation[k]{organization={Voronezh State University},  addressline={Universitetskaya square 1, Voronezh, 1394018, Russia}}

\begin{abstract}
Electron antineutrinos are emitted in the decay chains of the fission products inside a reactor core and could be used for remote monitoring of nuclear reactors. The DANSS detector is placed under the core of the 3.1~GW power reactor at the Kalinin Nuclear Power Plant (KNPP) and collects up to 5000 antineutrino events per day. DANSS measured changes of the reactor power by antineutrino counting rates over 7 years with 1.0\% accuracy in one week periods. The fission fractions of four major isotopes for the reactor power calculations were provided by KNPP. The systematic uncertainty of this measurement is less than 0.8\%. It is comparable to the accuracy of conventional methods of the reactor power measurements while it is based on completely different approach.
For the first time the \pu{39} and \ur{5} fission fractions were reconstructed  using antineutrino inverse beta-decay spectrum which is a completely new technique. This method was applied to the data from three reactor fuel campaigns (approximately 1.5 year each). The reconstructed fission fractions in about two-week measurements and the fission fractions provided by KNPP coincide within better than 3\% accuracy. This provides confidence in both completely different approaches of the fission fraction determination.
\end{abstract}

\begin{keyword}
%% keywords here, in the form: keyword \sep keyword, up to a maximum of 6 keywords
nuclear reactor, antineutrino, reactor monitoring, non-proliferation



\end{keyword}

\end{frontmatter}

%\tableofcontents



\section{Introduction}

The idea to use neutrinos to remotely monitor reactors was first proposed by Mikaelyan and collaborators~\cite{Mikaelyan:1978, Mikaelyan:1994}.
Electron antineutrinos are emitted in the decay chains of the fission products inside a reactor core.  They interact only through the weak force and therefore they pass through the containment building almost without attenuation. Neutrinos carry information about the reactor power and fuel content directly from the reactor core. Reactor antineutrinos are used in a wide range of fundamental studies that include the determination of the Pontecorvo-Maki-Nakagawa-Sakata (PMNS) matrix parameters (for a recent review see e.g. \cite{SajjadAthar:2021prg}), searches for sterile neutrinos(for a recent review see e.g. \cite{Danilov:2022str,Boser:2019rta,Coloma:2020ajw}), searches for the coherent neutrino scattering (for a recent review see e.g. \cite{Abdullah:2022zue}), searches for effects beyond Standard Model (SM) like a relatively large neutrino magnetic moment or Large Extra Dimensions (for a recent review see e.g. \cite{Acero:2022wqg}). For all these studies a precise knowledge of the antineutrino spectrum, fuel composition and a reactor power are very important. A possibility to determine remotely and non-intrusively the amount of \pu{39} produced during the reactor operation is interesting as an additional novel approach to the standard protocols for nuclear non-proliferation purposes (see e.g. \cite{Bernstein:2019hix,Carr:2018tak,NUCIFER:2015hdd}). A comparison between conventional methods for a reactor power determination and completely different neutrino based approach is interesting for understanding of the uncertainties in both methods and potentially could be used for the optimization of the reactor operations.

There are two approaches of the antineutrino spectrum predictions. In summation models all known decay chains are summed up (for recent models see e.g. \cite{Estienne:2019ujo,Letourneau:2022kfs,Perisse:2023efm,Vlasenko:2023eaf}). In conversion models the measured electron spectra of fission products of main isotopes \ur{5}, \pu{39}, \ur{8}, and  \pu{41} are converted into the antineutrino spectra. The most famous conversion model is the Hueber-Mueller (HM) model. It is based on the ILL measurements~\cite{SCHRECKENBACH1981251, SCHRECKENBACH1985325, VONFEILITZSCH1982162, HAHN1989365} of electron spectra for \ur{5}, \pu{39}, and \pu{41} fission products reconstructed by Huber~\cite{Huber}. The antineutrino spectrum of \ur{8} was calculated by Mueller and his coauthors in~\cite{Mueller}. The HM model is very often used as a benchmark model in many studies although it predicts about 6\% larger rate of antineutrino events than is detected using the inverse beta decay (IBD) reaction at different experiments (see e.g. \cite{Boser:2019rta}).
It also does not describe well the shape of the antineutrino spectrum. The experimental measurements are higher than the model predictions in the vicinity of 5 MeV prompt energy i.e. the total energy of a positron from the IBD process including the energy of photons from its annihilation~\cite{NEOS:2016wee, RENO:2020dxd, PROSPECT:2022wlf, DayaBay:2021owf, STEREO:2020hup, Stereo:2021wfd, DayaBay:2021dqj}. The discrepancy in the rate is probably caused by the overestimation in the model of the \ur{5} contribution.
The IBD yield for \ur{5} measured by the Daya Bay~\cite{DayaBay:2017jkb} and RENO~\cite{PhysRevLett.122.232501} experiments is smaller than the HM model predictions. Moreover, the recent measurements of the \ur{5} fraction in the electron yield by the Kurchatov Institute (KI) group~\cite{Kopeikin:2021rnb} are about 6\% smaller than the ILL results~\cite{SCHRECKENBACH1981251, SCHRECKENBACH1985325, VONFEILITZSCH1982162, HAHN1989365} that are used in the HM model. This can explain the difference in the rate. The conversion model based on the KI measurements of the \ur{5} contribution will be called the KI model~\cite{Kopeikin:2021ugh}. 
Recent summation models~\cite{Estienne:2019ujo,Letourneau:2022kfs,Perisse:2023efm,Vlasenko:2023eaf} predict IBD event rates consistent with the experimental measurements at industrial reactors. However they do not describe well the shape of the spectrum~\cite{DayaBay:2022jnn}. Conversion and summation models predict different shapes of the antineutrino spectrum~\cite{Perisse:2023efm}.

\begin{figure}[!htb]
\includegraphics[width=0.95\linewidth]{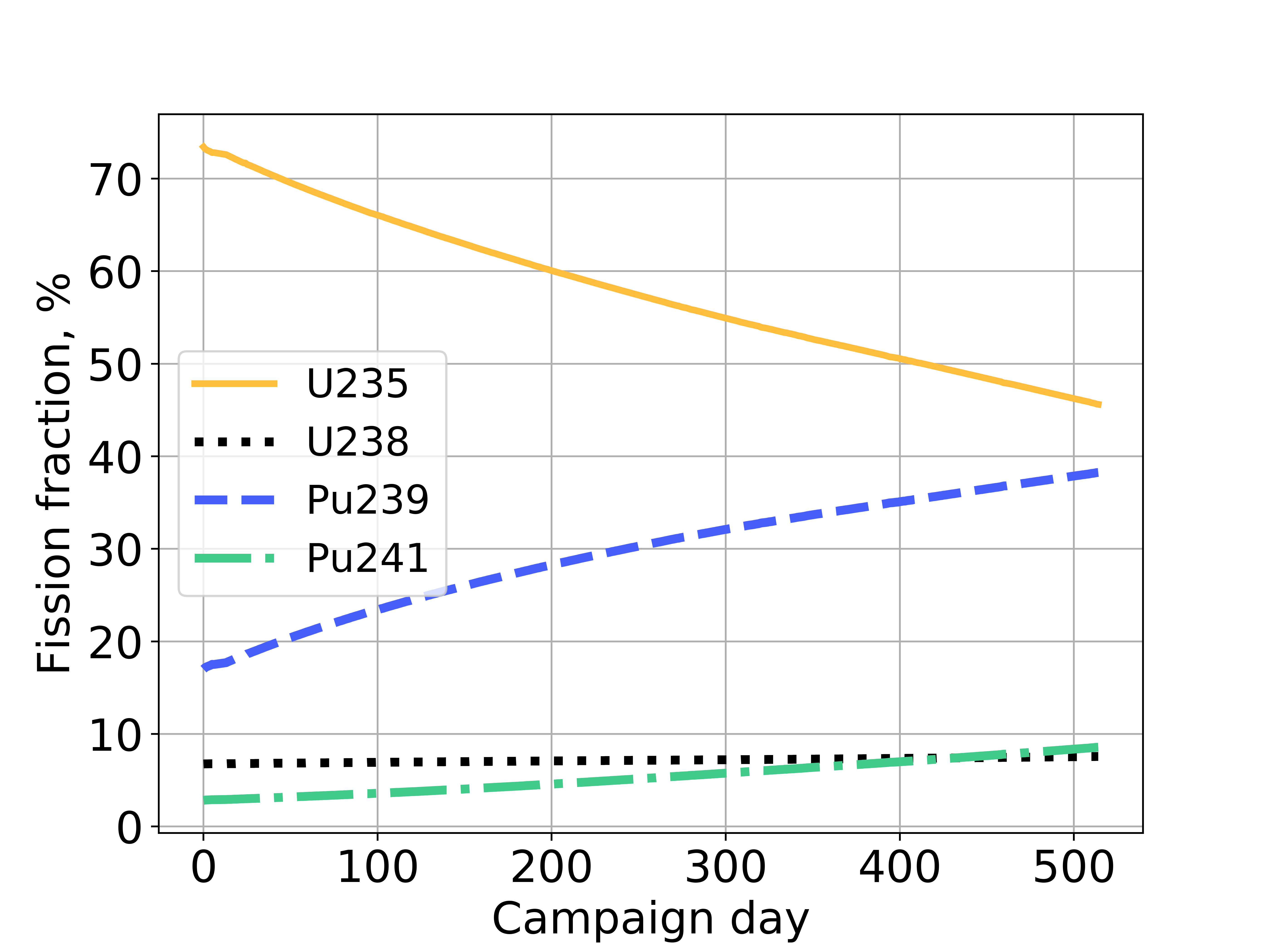}% Here is how to import EPS art
\caption{\label{fig:ff} A typical evolution of fission fractions with time of a power reactor operation (data are from KNPP).}
\end{figure}

In order to calculate the antineutrino spectrum one has to know the fission fractions of main isotopes i.e.  the ratio of the fission rate of a particular nuclide to the total fission rate. These fractions evolve with time during a reactor campaign. A typical fission fraction evolution during a fuel cycle for a power reactor is shown in Fig.~\ref{fig:ff}.
The fission fractions are usually estimated using very detailed and complicated simulations of neutron fluxes in reactors.
The accuracy in the fission fraction calculations is estimated by comparisons with measured fractions of different isotopes in a few fuel elements extracted from reactors after different exposures. The obtained spread between the measurements and predictions  is about $\delta f/f = 10\%$ for \ur{5} and \pu{39}, for different positions along the fuel pins (see~\cite{DayaBay:2017jkb,Barresi:2023sfw} and references therein).
After averaging over the sample position along the fuel pin (without edges) the agreement with predictions improves for \ur{5} and  \pu{39}, to about $\delta f/f = 5\%$.

In this paper we present a completely different approach based on a comparison of the measured antineutrino spectrum and model predictions for four main isotopes \ur{5}, \pu{39}, \ur{8}, and  \pu{41}.

\section{Kalinin NPP and DANSS detector}

The Kalinin Nuclear Power Plant (KNPP) in Russia consists of four power units with pressurized water power reactors (WWER-1000) with installed capacity of 1000~MW each. The thermal power is 3.1~GW for each unit.%\SI{1000}{\mega\watt}
%Installed capacity of each unit is 1000~MW and the thermal power is 3.1~GW.

Only one third of new fuel assemblies are added before the start of a new reactor campaign.  Remaining two thirds are reshuffled in order to maximize the energy output from the fuel~\cite{Zhutikov:2024fhi}. Some (64 at KNPP)
fuel assemblies in the WWER-1000 reactors are equipped  with 7 direct charge detectors placed at different positions along the fuel assembly. 
These  detectors measure the charge induced by beta decays  and hence the local released power in several hundred points distributed in the reactor active zone. The spatial distribution of the released power changes during the reactor campaign both along the reactor radius and the reactor height~\cite{Zhutikov:2024fhi}. However these changes are not large and they are not corrected for in the present analysis in order to make it independent from the information from KNPP. 
The energy release distribution and burnup depth are calculated for all fuel assemblies (163) inside the reactor core using the BIPR-7A program (which is a part of the CASCADE software~\cite{CASCADE}). These calculations are verified by comparison between the predicted and measured power distribution along the fuel assembly using direct charge detectors. The predicted power distribution agrees all the time with the measured one (the standard deviation of the difference between the calculated values of power and the in-core monitoring system data is $\sim2.8\%$~\cite{Zhutikov:2024fhi}).
The BIPR-7A program is a coarse-mesh deterministic code, which allows for  
fast calculations up to a real-time operation. It solves the diffusion equation for only 2 groups of neutron energy using effective parametrization of necessary parameters, such as the neutron multiplication coefficient, migration area, etc., obtained in a more detailed TVS-M program. 
The BIPR-7A package uses prism centers as calculation nodes, into which fuel assemblies are conventionally divided by height (usually into 16 layers). The BIPR-7A program performs calculations and provides information about the three-dimensional distribution of the following main parameters: fuel burnup; temperature and density of the coolant; power and neutron fluxes; concentrations of $^{135}$Xe, $^{149}$Sm, and $^{149}$Pm isotopes.
The deterministic TVS-M~\cite{articleTVS-M, TVSM1, TVSM2} program uses a detailed geometry of the fuel assemblies and 69 neutron energy groups. It is similar to the widely known WIMS program~\cite{Askew1966GENERALDO, wims9}. The results of the TVS-M program are verified~\cite{TVS-MvsMCU} using a Monte Carlo MCU program~\cite{MCU} with a very detailed description of the reactor core geometry and parameters.  

The TVS-M program is used for the transition from the energy release and burnup depth to the distribution of the fission fractions.  
The expected accuracy in fission fractions is about $\delta f/f =$ 5\% (according to KNPP personnel).

\ur{5} and \pu{39} are the main fission nuclides. The \ur{5} fission fraction decreases roughly from 73\% to 45\%. The \pu{39} fission fraction increases from 17\% to 38\%. The
\ur{8} fission fraction changes from 7\% to 8\% during a typical fuel cycle and the \pu{41} fission fraction changes from 3\% to 9\%. Their contribution is small but not negligible. 
Different nuclides have different \antinu\ spectra (both in the absolute normalization and shape), therefore an \antinu\ detector is sensitive to the fuel evolution.

DANSS (Detector AntiNeutrino based on Solid Scintillator)~\cite{DANSS:design} is located at the unit 4 of KNPP
under the reactor core. DANSS is placed on a moving platform which allows to vary the distance from the detector to the reactor core from 10.9 to 12.9~m (center to center). The reactor building provides about $\sim$50~m~w.e. overburden which practically removes the hadronic component of cosmic rays and decreases the muon flux by a factor of 6. DANSS consists of 2500  plastic scintillator counters ($100\times 1\times 4$~cm$^3$ each)  with wave length  shifting fibers read out with Silicon Photo-multipliers and usual photo-multipliers. The total sensitive volume is one cubic meter. The detector is surrounded by several layers of passive and active shielding.
The inverse beta decay (IBD) reaction is used to detect antineutrino: $\antiparticle\nu_e + p \to n + e^+$, with the positron energy $E_{e^+} \approx E_{\antiparticle\nu} - 1.8~\mathrm{MeV}$. The IBD event consists of two signals separated by (1--50) microseconds - the prompt and delayed signals. The prompt signal is produced by a positron and its annihilation. The prompt signal carries the information about the antineutrino energy. A high DANSS   segmentation allows to measure a positron %pure 
kinetic  energy without the  energy of the annihilation photons. The neutron after thermalization is captured by gadolinium. About 8~MeV gamma burst from the Gd deexitation gives a delayed signal. The IBD event selection criteria are described in~\cite{DANSS:PLB18,Svirida:2020zpk}.   The accidental background of about 23\% of the IBD rate is measured directly by the experiment in sixteen 49 $\mathrm{\mu s}$ time intervals preceding the neutron candidate by 5, 10, ..., 80~ms. In the analysis coincidences  between the prompt positron signal and delayed neutron signal of the IBD event are searched for in a 49 $\mathrm{\mu s}$ window. At larger times the probability to observe the neutron candidate decreases exponentially. Therefore at time intervals larger than 5~ms only accidental coincidences are possible. Averaging the accidental coincidence rate in 16 time intervals improves its accuracy by a factor of 4. In such a way the accidental background is subtracted from the data in a completely model independent way with a high precision.
The correlated background produced mainly by the cosmic muons is only 1.8\% of the IBD signal in the detector position closest to the reactor where DANSS detects more than 5000 IBD events per day. The shape of this background is determined using signals tagged with the muon VETO system. Normalization is determined using reactor off periods.
More than 8 million IBD events with the very small (and subtracted) background were collected by DANSS during 7 years of operation. This is the largest sample of neutrino events among all reactor neutrino experiments. One of the goals of the  DANSS project is  a search for sterile neutrinos.
The latest results are presented in~\cite{Alekseev:2024mkb}.

\section{Measurements of reactor power with neutrinos}

\begin{figure*}[!htb]

\includegraphics[width=0.95\textwidth]{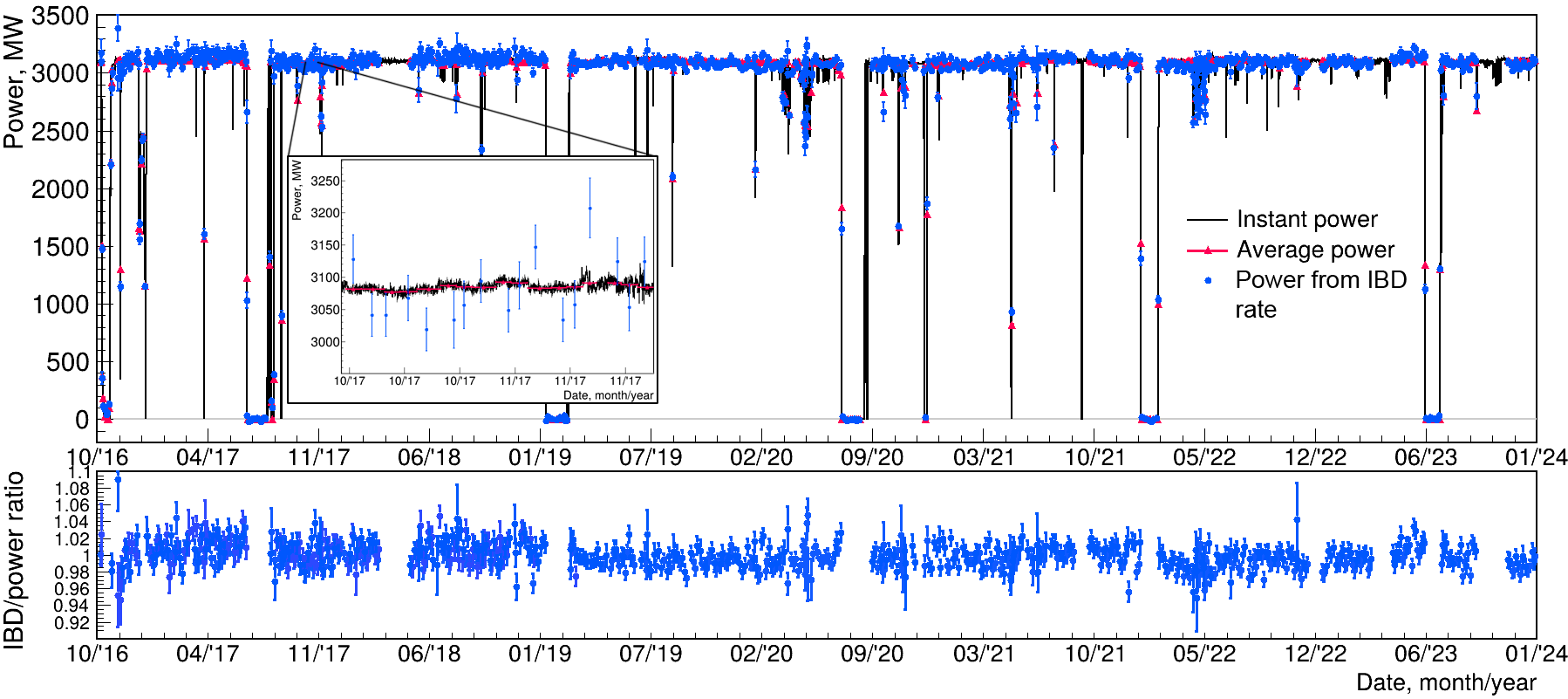}
\caption{\label{fig:power}Top panel: KNPP power measured with conventional methods (black line -- instant power, red triangles -- average power) and with the IBD rate at 3 different detector positions (blue circles) corrected for the variations in the detector efficiency, fuel evolution, dead time, and the different distances from the detector to the reactor core. The IBD rate was normalized to the reactor thermal power during one month in 2016. Bottom panel: ratio of the two measurements.  Enlarged picture of about one month period is shown in the
insertion.}
\end{figure*}

The measurement  of a nuclear reactor  power is a very important part of
its operation process. 
Conventional techniques are well developed and provide the accuracy of about 0.8\% according to the information from KNPP personnel.
In order to achieve such a high  accuracy many systematic uncertainties were carefully estimated which is by far a non-trivial task.
Therefore the development of independent methods of the power measurement based on alternative physics principles is very important for the improvement of the safety and reliability of nuclear reactors.
The possibility to use antineutrino detectors for reactor monitoring  was proposed long time ago~\cite{Mikaelyan:1978, Mikaelyan:1994}.
The spectrum of the reactor antineutrinos with energy $E_{\nu}$ detected at time $t$ is
%\begin{fleqn}
\begin{equation}
\label{eq:spectrum}
\frac{d^2N}{d E_{\nu}  dt} =  N_p \sigma (E_{\nu}) \epsilon(E_{\nu}) \frac{d^2\phi(E_{\nu}, t)}{dE_{\nu}dt} \frac{1}{4\pi L^2} P(E_{\nu}, L),
\end{equation}
%\end{fleqn}
where $N_p$ is the number of target protons, $\sigma(E_{\nu} )$ is the IBD reaction cross section, $\epsilon(E_{\nu})$ is the 
detector efficiency, $L$ is the distance between the centers of the detector and reactor core (the detector and reactor sizes and distribution of fission points are taken into account by appropriate integration everywhere in the article where it is not stated otherwise), and $P(E_{\nu}, L)$  is the survival probability due to neutrino oscillations (assumed to be 1 in this paper). The term $d^2\phi(E_{\nu}, t)/dE_{\nu} dt$ is the antineutrino spectrum and
\begin{equation}
\label{eq:flux}
\phi(E_{\nu}, t) = \frac{W_{th} (t)}{\overline{E_{fiss}}(t)} \sum f_i(t)\cdot s_i(E_{\nu})
,
\end{equation}
where the index $i$ runs over the four primary fission isotopes
(\ur{5}, \pu{39}, \ur{8}, and  \pu{41}), $W_{th}(t)$ is the reactor thermal power, $f_i (t)$ is the fission fraction of isotope $i$, $\overline{E_{fiss}}(t) = \sum f_i (t) e_i$ is the average energy release per fission due to the energy release
$e_i$~\cite{PhysRevC.88.014605} from each fission isotope, and $s_i (E_{\nu} )$ is the antineutrino energy spectrum per fission for individual isotopes.

\begin{figure}[!htb]
\includegraphics[width=0.85\linewidth]{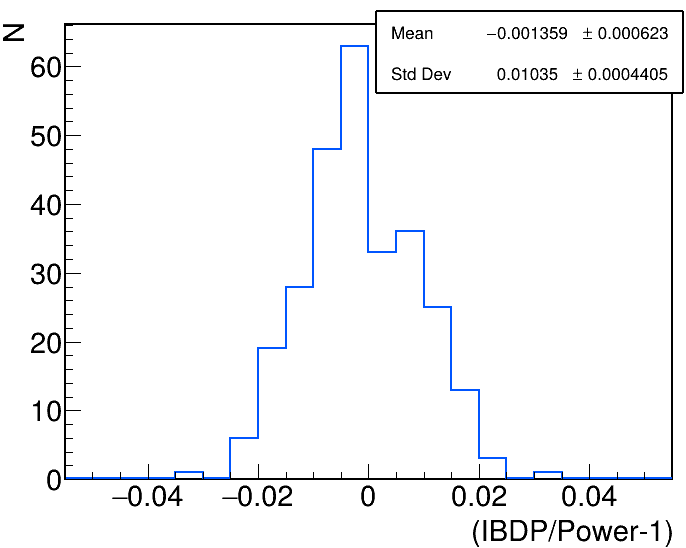}
\caption{\label{fig:power-difference} Deviation from unity of the ratio of the reactor power measured with IBD rate (IBDP) and with the conventional methods for 7 years of the reactor operation. }
\end{figure}

Unfortunately, uncertainties in the reactor model predictions for the antineutrino 
spectra per fission $s_i(E_{\nu})$ are at present too big (about 5\%) to use them for the absolute antineutrino flux predictions and hence for a reactor power determination with a competitive uncertainties in comparison with the conventional methods.
Therefore relative measurements of the antineutrino flux were used in this paper for the reactor power determination.
The IBD event rate during one month in November--December 2016 was normalized to the reactor power provided by KNPP.
This allows to express the IBD counting rates in units of the reactor power (MW). After that during 7 years the measured IBD rate was corrected for the changes in the fuel composition during the reactor campaigns
using the HM model for antineutrino spectra from the fission products of 4 main isotopes \ur{5}, \pu{39}, \ur{8}, and  \pu{41}. The fission fractions for these isotopes were provided by KNPP.  The measured IBD event rate was also corrected  for small changes of the detector efficiency and dead time. The changes of the efficiency of less than 3\% are mainly due to appearance and disappearance of dead channels and they are not correlated with the detector position. The changes of the dead time (about 1\%) are correlated with the muon flux and hence with the detector position. However they are directly corrected using the measured muon VETO counting rate.
After these corrections the agreement between the reactor power measured with conventional methods and measured with neutrinos becomes excellent (see Fig.~\ref{fig:power}). The time periods for each data point are determined mainly by the detector movement between different positions. For the first data taking phase (till December 2018) with 3 detector positions the periods were about 2--3 days and the corresponding statistical accuracy of the points was about 1.2\%. For the second data taking phase the time periods for each point were about 1 week and the corresponding statistical accuracy was about 0.7\%.
To obtain a quantitative estimate of the accuracy the reactor off periods were excluded by the requirement on the power to be larger than 60\% of the maximum power. 
Data from Fig.2 were combined into about one week periods.
The difference between the two methods of the power determination for about one week measurement periods has a spread  of $\sigma$=1.0\%  and the mean value consistent with zero (-0.14\%) for the whole period of 7 years (see Fig.~\ref{fig:power-difference}). This spread is  larger than the average statistical error of the measurements of 0.67\%.
This can be interpreted as the evidence for additional systematic error of about 0.79\% obtained by subtracting in quadrature the measured statistical error of  0.67\% from the total spread of 1.04\%. 
The systematic error includes both the systematic uncertainties in the DANSS measurements as well as possible systematic uncertainties in the conventional methods of the reactor power determination.
Thus the obtained accuracy in the remote relative reactor power determination using neutrinos is comparable to the accuracy of the conventional methods based on completely different physical principles.  This provides additional 
information for estimates of systematic uncertainties of the conventional methods for the reactor power determination.
The systematic uncertainty of 0.79\% does not include a possible bias on the absolute normalization from conventional methods, but only additional uncertainties that can affect the conventional power measurement on a relative scale.

\section{Determination of fission fractions using antineutrinos}

The IBD positron energy spectrum is proportional to a sum of individual isotope spectra weighted with the corresponding fission fractions. If we take equations (\ref{eq:spectrum}--\ref{eq:flux}), express $E_{\nu}$ in terms of $E_{e+}$, and keep terms which depend only on $f_i$, the IBD spectrum is described by
\begin{fleqn}
\begin{equation}
\label{eq:spec_positron}
\frac{dN}{d E_{e+} } =  \int_{E_{\nu}} \alpha  W_{th} \sigma (E_{\nu}) \epsilon(E_{\nu}, E_{e+}) \frac{\sum f_i(t)\cdot s_i(E_{\nu})}{\sum f_i (t) e_i}  dE_{\nu},
\end{equation}
\end{fleqn}
where $\alpha$ includes the reactor and  detector geometry, the number of protons in the detector, and other factors independent of the fission fraction changes throughout the campaign, and $\epsilon(E_{\nu}, E_{e+})$ describes the detector response. 
The DANSS response is calculated using Monte~Carlo (MC) simulations performed with the Geant4 package.
In this section we consider measurements performed at full reactor power, so $W_{th}$ remains almost constant (within $\sim1\%$) in this formula.
Therefore the fission fractions can be determined from the fit of the measured spectrum if individual energy spectra are known.

\begin{figure}[!h] %!htb
\includegraphics[width=0.89\linewidth]{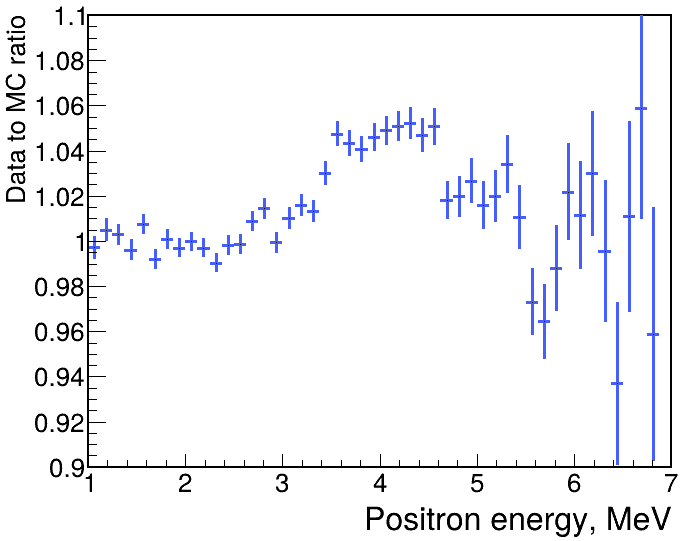}
\caption{\label{fig:bump} Data to the HM model based MC ratio for the positron energy spectrum collected during campaign 5 (statistical errors only).
}
\end{figure}

The HM model was used for the spectra predictions for individual isotopes. Since reactor models do not describe well the IBD prompt signal spectrum around 5~MeV the positron energies from 3 to 5.5~MeV were excluded from the fit (the positron kinetic energy is about 1 MeV smaller than the prompt energy). 
The ratio of the observed positron spectrum to the predicted one is presented in Fig.~\ref{fig:bump} for the \fth campaign. The predicted positron spectrum is calculated using MC simulations with the embedded antineutrino spectrum from the HM model.

For this campaign used to fix the energy scale and shift parameters fission fraction values were fixed to the KNPP data.
The DANSS energy scale is determined mainly from the measurements of the $^{12}$B beta spectra cross-checked with measurements of several radioactive sources~\cite{Svirida:2020zpk}. The scale uncertainty is conservatively  estimated as  2\%. In addition a possible shift in energy is considered with a conservative uncertainty of 50~keV.
These parameters were determined during the fit of the HM model predictions to the data of the \fth\ campaign. The fit ranges were  (1--3)~MeV and (5.5--7)~MeV.
The fit is performed by minimization of $\chi^2$:
\begin{equation}
\label{eq^chi2}
\chi^2 = \sum_{i}\frac{(N_i^{\mathrm{obs}} - N_i^{\mathrm{pre}}(\mathbf{\eta}))^2}{\sigma_i^2},    
\end{equation}
where $N^{\mathrm{obs}} (N^{\mathrm{pre}})$ -- observed (predicted) counts in bin $i$, $\mathbf{\eta}$ -- parameter vector (energy scale and shift), $\sigma$ -- statistical error in each bin.
The fit quality is very modest (p-value is 1\%) but acceptable. 
The obtained value of the energy shift is ($57 \pm 5$)~keV while the energy scale is consistent with the default value (the ratio is $1.001\pm 0.001$). In further fits of positron spectra we use obtained values of the energy shift and scale.

The \ur{8} and  \pu{41} fission fractions change very little during the reactor campaigns (see Fig.~\ref{fig:ff}).
Hence their dependence on the number of days of the reactor operation was parameterized by polynomial functions using the KNPP data on the \ur{8} and  \pu{41} fission fractions for the first full reactor campaign observed by  DANSS (campaign 5). 
Therefore we fit only the fission fraction of  \pu{39} (since the sum of the fission fractions is equal to 1).

\begin{figure*}[!htb]
\includegraphics[width=0.98\linewidth]{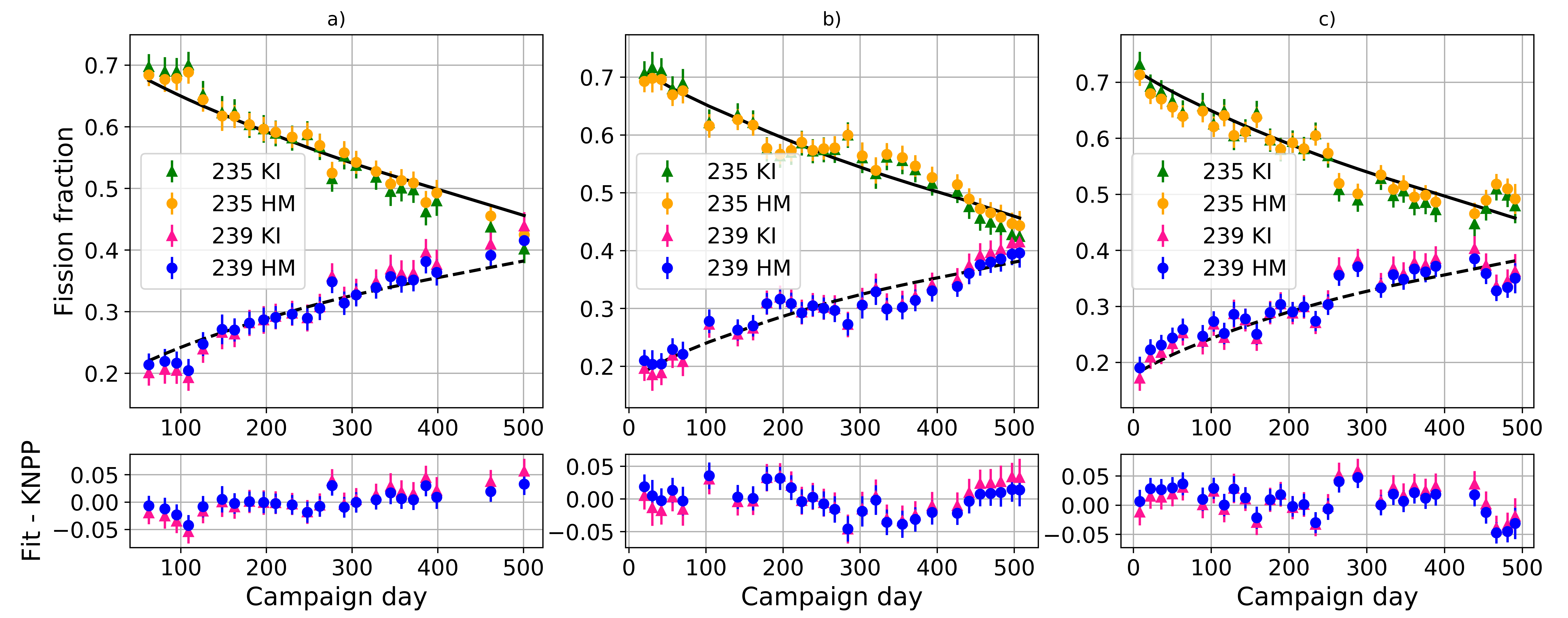}
\caption{\label{fig:fit_result} 
Top: Fission fractions for \ur{5} and \pu{39} obtained from the fit of the antineutrino IBD spectrum using the HM model (yellow and blue circles correspondingly) and  the KI model (green and red triangles correspondingly). Data from KNPP are shown with solid and dashed lines correspondingly. Bottom: Difference between the \pu{39} fission fractions obtained  from KNPP and from the fits of the IBD spectrum using the HM (blue) and KI (red) models. Results for campaigns 6, 7, and 8 are presented in a), b), and c) panels correspondingly.
}
\end{figure*}

Measurements made at the top (closest to the reactor) and  bottom detector positions were combined together with a correction for different distances. 
Each data point corresponds to about two weeks of data taking and has a statistical error in the number of IBD events of about 0.6\%. However there are several points with shorter data taking periods because of the problems with the detector, reactor, and/or infrastructure.
During the fit procedure the mean normalization for the whole considered campaign is used  in order to be independent from the uncertainties in the absolute rate predictions and the absolute detector efficiency as well as from the average reactor power during a corresponding campaign.
This means that after observing the full reactor campaign the observed to predicted counting rate ratio was calculated using the model predictions for  mean fission fractions provided by KNPP for the \fth\ campaign. 
The actual fission fractions provided by KNPP for the studied campaign are not used in the normalization.
In the following fits the model predictions were multiplied by the obtained ratio.
Corrections for the dead time, efficiency, and contributions from the neighbor reactors (about 0.6\%)  were done for each individual measurement. 
In contrast, the reactor  power and fission points distribution profile were  not taken into account in order to make the method more independent from the reactor information.
Fig.~\ref{fig:fit_result} shows the \pu{39} and \ur{35} fission fractions obtained from the fit of the HM model to the data from campaigns 6--8.
The fit quality of individual data periods is quite good (average $\chi^2=28$ for 25 degrees of freedom).
The best fit values for the nuisance parameters  from the \fth\ campaign were used in the fits for studied campaigns 6--8.

We don’t introduce uncertainties related to the HM model in our fit. They are not known well enough. We treat the HM model as an effective description of the antineutrino spectra for different isotopes. All imperfections of this description influence the spread between the fission fractions provided by KNPP and those obtained from the fit of the positron spectra. This spread includes also all other possible systematic uncertainties including uncertainties from the detector performance, uncertainties arising from reactor instabilities  as well as the uncertainties in the fission fraction predictions provided by KNPP which are based on the reactor simulation programs.
The average error on the \pu{39} fission fraction obtained from the fits is $1.9\%$ per point. It can be considered as the statistical error since the model for the IBD positron spectra is fixed in the fit and does not contain any systematic uncertainties.

\begin{figure}[!hb]
\includegraphics[width=0.84\linewidth]{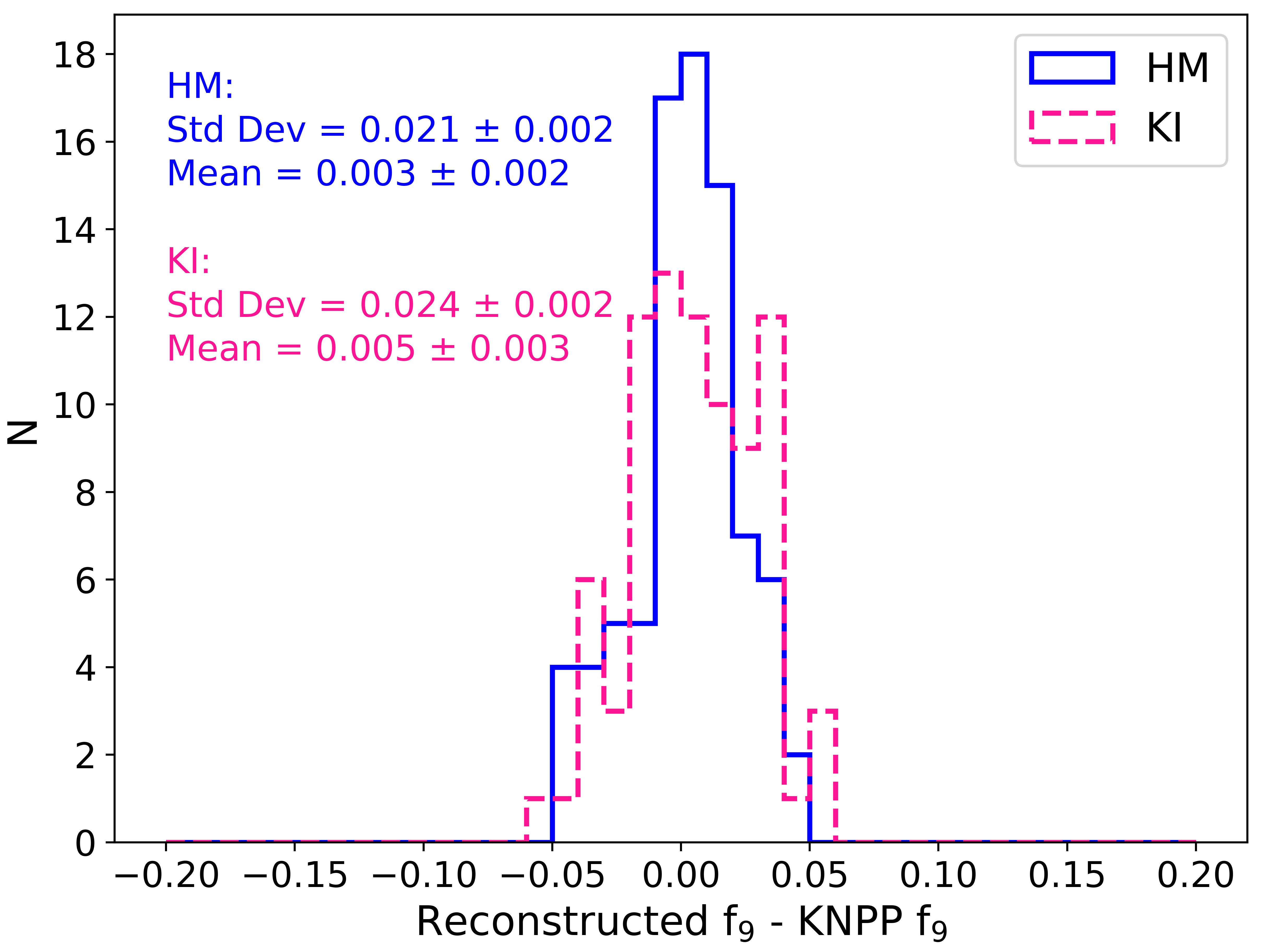}
\caption{\label{fig:deviation} 
Difference between the  \pu{39} fission fractions obtained from the fit of the positron energy spectrum using the HM (solid histogram) and KI (dashed histogram) models and from the KNPP data for campaigns 6–8.
}
\end{figure}

The agreement between the DANSS and KNPP results is very good. The difference  between the fit results and the KNPP fission fractions  (reconstructed $f_{239}$ - KNPP $f_{239}$) for campaigns 6--8 used in the analysis has the standard deviation of 2.1$\pm 0.2\%$ and the mean value of  $0.3\pm 0.2\%$ consistent with zero (see Fig.~\ref{fig:deviation}). The relative difference ((reconstructed $f_{239}$ - KNPP $f_{239}$)/KNPP $f_{239})$ decreases with the rise of $f_{239}$ and  has on average the standard deviation of about 7\%.
The campaign 5 is not considered here because the corresponding data were used for the parametrization of the \ur{8} and \pu{41} fission fractions dependence on time. The excellent agreement between the two fission fraction determinations based on completely different approaches and different data provides confidence in the validity of both approaches. 
The analysis was performed also using the KI model~\cite{Kopeikin:2021rnb, Kopeikin:2021ugh} for the antineutrino spectra. 
The results are shown in Figures \ref{fig:fit_result} and \ref{fig:deviation}.
The standard deviation of the difference between the fit results 
and KNPP data on the fission fractions
of 2.4$\pm 0.2\%$ (see Fig.~\ref{fig:deviation}) is similar to the result obtained  with the HM model and the mean value of $0.5\pm 0.3\%$ is consistent with zero. 
Main information about fission fractions is related to the IBD rate. The KI model has about 5\% smaller IBD yield caused by \ur{35}. Therefore the use of the KI model leads to a systematic shift in the obtained fission fractions in comparison with the HM model as seen in Fig.~\ref{fig:fit_result}. At the beginning of the campaign the KI model gives smaller fractions of \pu{39} while at the end of the campaign it gives larger fractions of \pu{39}. However the difference with the fission fractions provided by  KNPP is still reasonably small.

\section{Conclusions}

The most comprehensive results on the reactor monitoring using antineutrinos are presented.
The relative reactor power was measured remotely using antineutrinos with 1.0\% accuracy in one week measurements during 7 years of the reactor operation using the fission fractions of four major isotopes provided by KNPP. The absolute reactor power is not measured with this method since the absolute antineutrino event rate is not used in the analysis due to its large uncertainties.
The systematic uncertainty was estimated by subtracting in quadrature the statistical errors from the total measured errors. It was found to be smaller than 0.8\%. This systematic uncertainty includes both the systematic uncertainties in the DANSS measurements as well as possible systematic uncertainties in the conventional methods of the power determination. Thus the measurements of the reactor power using neutrinos agree very precisely with the conventional methods although they use completely different physics processes and have completely different sources of systematic uncertainties.
This provides confidence in the estimates of the uncertainties in the reactor power determination using conventional methods.

The \ur{5} and  \pu{39} fission fractions were determined from the fit of the measured antineutrino IBD spectrum using reactor model predictions for the spectra of four main isotopes \ur{5}, \pu{39}, \ur{8}, and  \pu{41}.  They agree with better than 3\% accuracy in the absolute \pu{39} fission fractions with the fission fractions provided by KNPP which are based on the neutron flux simulations inside the reactor. The two approaches use completely different data and equipment and hence they have completely different systematic uncertainties. The excellent agreement between the two approaches provides confidence in their validity.

The obtained results are important for the numerous precision studies in fundamental physics at reactors  since they help to constrain the antineutrino spectrum. They could be also interesting for the nuclear weapon non\-/proliferation control and for improvements in the reactor operation.

\section*{Acknowledgments}
The DANSS collaboration deeply values the permanent assistance and help provided by the administration and staff of KNPP.
This work is supported in the framework of the State project Science by the Ministry of Science and Higher Education of the Russian Federation, Grant No. 075-15-2024-541.





%% For example:

%% \bibitem[Aladro et al.(2015)]{Aladro15} Aladro, R., Martín, S., Riquelme, D., et al. 2015, \aas, 579, A101

%%\end{thebibliography}

\end{document}